\documentclass[runningheads]{llncs}
\usepackage[T1]{fontenc}
\usepackage{amsmath,amssymb,amsfonts}
\usepackage{mathtools}
\usepackage{graphicx}
\usepackage{algpseudocode}
\usepackage{algorithm}
\usepackage[caption=true]{subfig}
\usepackage{hyperref}
\usepackage{adjustbox}
\usepackage{color}

\begin{document}
\title{Trace-based, time-resolved analysis of MPI application performance using standard metrics}
\titlerunning{Time-resolved MPI performance metrics}
%
\author{Kingshuk Haldar\inst{1}\orcidID{0000-0001-9222-7488}}
\authorrunning{K. Haldar}
%
\institute{{High Performance Computing Center Stuttgart (HLRS), University of Stuttgart}
\email{{haldar.kingshuk@gmail.com}, {kingshuk.haldar@hlrs.de}}}
\maketitle              
\begin{abstract}
Detailed trace analysis of MPI applications is essential for performance engineering, but growing trace sizes and complex communication behaviour often render comprehensive visual inspection impractical.
This work presents a trace-based calculation of time-resolved values of standard MPI performance metrics, load balance, serialisation, and transfer efficiency, by discretising execution traces into fixed or adaptive time segments.
The implementation processes Paraver traces post-mortem, reconstructing critical execution paths and handling common event anomalies, such as clock inconsistencies and unmatched MPI events, to robustly calculate metrics for each segment.
The calculated per-window metric values expose transient performance bottlenecks that the time-aggregated metrics from existing tools may conceal.
Evaluations on a synthetic benchmark and real-world applications (\texttt{LaMEM} and \texttt{ls1-MarDyn}) demonstrate how time-resolved metrics reveal localised performance bottlenecks obscured by global aggregates, offering a lightweight and scalable alternative even when trace visualisation is impractical.

\end{abstract}

\section{Introduction}
\label{sec:intro}
The size of trace files commonly increases on larger systems because more MPI processes result in more MPI calls, which are directly proportional to the trace file size.
Analysing large traces remains challenging as it requires specialised techniques, domain expertise, and considerable effort.
Simpler metrics, such as speed-up, bypass the challenges of handling larger traces and still highlight potential scaling inefficiencies; they cannot offer further meaningful insights into the root causes of bottlenecks or optimisation opportunities.
Rosas et al.~\cite{modelfactors14} introduced the fundamental parallel performance factors, load balance, serialisation efficiency, and transfer efficiency, to quantify the contribution of different bottlenecks to suboptimal scaling.
Performance metrics provide a high-level summary of overall performance characteristics across the entire execution.
But they do not capture the transient behaviours caused by modules executed at different times and shifting computational loads.

The sheer volume and intricate communication patterns of large trace files may obscure the root cause of performance bottlenecks, as manual visual inspection in such situations is time-consuming and prone to misinterpretation.
Meanwhile, performance metrics calculated from recorded traces aggregate across the entire execution, masking local effects.
For instance, an iterative solver periodically performs serial I/O, temporarily stalling the other MPI ranks.
This local situation may be hidden when metrics are aggregated across the entire execution.

Traceless or online frameworks provide similar metrics without traces, bypassing the analysis overhead.
They have similar limitations of aggregated metrics and require re-running the application with the tools for any new analysis.
This step is not always practical or feasible in production environments.

\subsection{Contribution}
\begin{figure}[tp]
  \centering
  \includegraphics[width=0.85\linewidth]{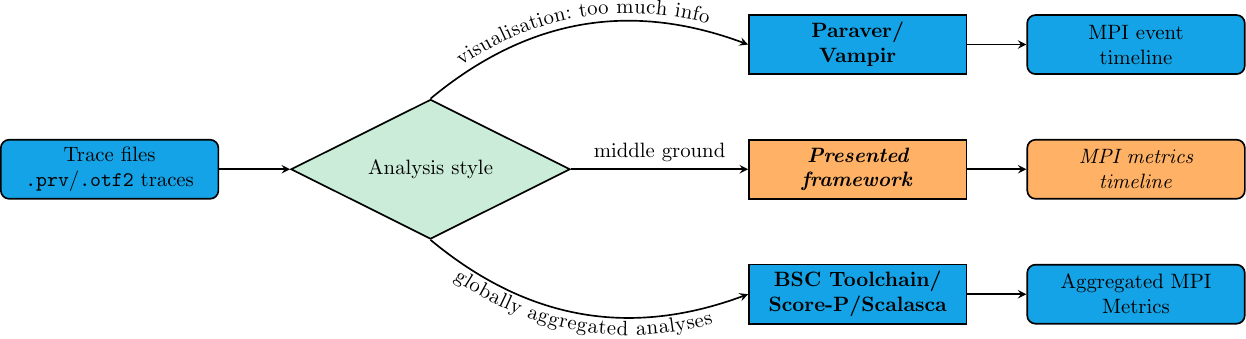}
 \caption{Conceptual placement of the framework within existing MPI performance analysis workflows. It bridges the gap between exhaustive trace visualisation and globally aggregated scalar metrics by enabling time-resolved metric views. These time-series plots help analysts identify localised inefficiencies that would otherwise be hidden in global values.}
 \label{fig:method-intro}
\end{figure}
Fig.~\ref{fig:method-intro} shows that analysts can arrive at different insights from the same trace file depending on the toolset/methodology choice.
The presented framework is placed in the middle to emphasise the combination of visualisation and quantitative insights it provides.
This work presents a trace-based framework for calculating time-resolved values of established MPI performance metrics.
Partitioning the execution timeline into discrete windows enables the framework to observe the localised performance and its evolution over time.

The key contributions are the following.
\newline
\textbf{Time-resolving framework} to compute existing performance metrics from recorded MPI traces over discretised time windows.
Using the same metrics as used in the standard globally aggregated calculation, this framework allows analysts to interpret them similarly to existing frameworks.
At the same time, those values are also duration-aware for the designated time window without requiring manual instrumentation or repeated application runs.
\newline
\textbf{Discretisation lengths' effects} are explored for practical application of the framework.
The investigated aspects include event sparsity, the effect of the critical path on clock consistency, and the visual stability of the metrics across various discretisation lengths.
\newline
\textbf{An implementation} is presented as a tool~\cite{clktk} which applies this framework to Paraver trace files~\cite{paraver}.
This visualisation complements, rather than replaces, traditional trace viewers.
The tool also performs the critical path analysis as part of the calculation.
\newline
\textbf{Evaluation} across synthetic and real-world traces demonstrates the effectiveness of localised metrics calculations in exposing performance issues that would otherwise remain concealed by globally aggregated metrics.

This work's contribution is not a new metric or performance model, but a robust, user-configurable framework to automate the time-localised computation of established MPI metrics post-mortem, which current tools do not efficiently offer.
This work complements the existing tools by offering a lightweight, post-mortem mechanism for time-localised analysis of MPI application performance.

\subsection{Organisation}
The paper is arranged as follows. Section~\ref{sec:background} provides a brief background of this work, Section~\ref{sec:discrete-framework} discusses discretisation framework for time-resolved performance metrics calculation, Section~\ref{sec:tool-des} elaborates the implementation details of the framework in the presented tool, the case-studies are reported in Section~\ref{sec:eval}, the related work is discussed in Section~{\ref{sec:rel-work}}, and Section~{\ref{sec:summary}} summarises the presented framework, observation from the case studies and the future outlook.

\section{Background}
\label{sec:background}
Performance metrics quantify the performance behaviour, which is important for systematic scaling studies.
Rosas et al. \cite{modelfactors14} introduced the fundamental performance factors to decompose the parallel efficiency of HPC applications into three components: load balance, serialisation, and transfer efficiency.
\begin{equation}
\label{eqn/pe}
\mathit{\eta_{||}}= \underbrace{\frac{{\sum\limits_i(\prescript{}{}{t_{i}^{compute}})} / P}{max(\prescript{}{}{t_{i}^{compute}})}}_\text{Load~Balance (LB)} \times \underbrace{\frac{max(\prescript{}{}{t_{i}^{compute}})}{runtime_{ideal~n/w}}}_\text{Serialisation (Ser)} \times
\underbrace{\frac{runtime_{ideal~n/w}}{runtime_{observed}}}_\text{Transfer (Trf)}
\end{equation}
$\mathit{\prescript{}{}{t_{i}^{compute}}}$ is the cumulative value of the out-of-MPI (OOM) durations of rank $\mathit{i}$, and $\mathit{max(\prescript{}{}{t_{i}^{compute}})}$ is its highest value among all $P$ processes with $\mathit{i}= \mathit{\{0, ..., P-1\}}$; and $t_{ideal~n/w}$ is obtained by calculating the critical execution path~\cite{critpath05} of the MPI application.
$\mathit{t_{observed}}$ is the recorded execution time of the application.
These multiplicative metrics quantify the impact of workload distribution, communication overlap, and communication costs on the overall parallel efficiency.
They are routinely used to conduct post-mortem performance analysis of applications across their execution.

Traces contain discrete MPI events with timestamps and related information about the associated point-to-point messages (such as tags or sizes) and collective communications (such as communicators, sizes, or roots).
Instrumenting an application with tracing tools~\cite{tools22} generates event-based traces, whose standard formats include Paraver and OTF-2.
For Paraver traces, the BSC toolchain's Basic Analysis~\cite{basicanalysis} calculates the fundamental performance factors.
For the OTF-2 format, similar tools are Scalasca~\cite{scalasca10}-based Cube and Scout.
Scout performs the critical path calculation at the \texttt{MPI\_Finalize} for root-cause analysis~\cite{scalascaroot16}.

Modern online and traceless approaches, such as the OTF-CPT~\cite{otf22}, track the critical path length and other quantities from~\eqref{eqn/pe} during the application's execution, eliminating the challenges of handling large traces and providing quick, lightweight performance assessment.

The existing tools compute standard metrics for entire executions, missing local details, or avoid trace collection entirely, sacrificing post-mortem flexibility.
This work aims to bridge this gap by calculating the established metrics in a time-resolved framework from recorded MPI traces, without requiring re-running or instrumentation.

\section{Time-Resolved Metric Computation Framework}
\label{sec:discrete-framework}
This section describes the framework for computing time-resolved performance metrics of MPI traces.
The framework performs post-mortem localised performance analysis using the performance metrics in~\eqref{eqn/pe}.

Since applications perform various tasks throughout execution, the corresponding performance metrics across different trace regions may differ.
This dissimilarity becomes evident when a specific time segment of recorded traces is analysed separately, and the resulting metrics exhibit performance characteristics that differ from those of the entire execution.
The presented framework automates this isolation and analysis of time segments with incremental time progression.
This results in consecutive time segments large enough for the metrics calculation but small enough to capture the localised behaviour.

\subsection{Timeline discretisation}
\begin{figure}[tp]
\begin{center}
\begin{minipage}[c]{\linewidth}
  \subfloat[Timeline and its equidistant discretisation. The blue and orange sections represent OOM and MPI regions, respectively.\label{fig:disc-cartoon}]
  {\includegraphics[width=.95\linewidth]{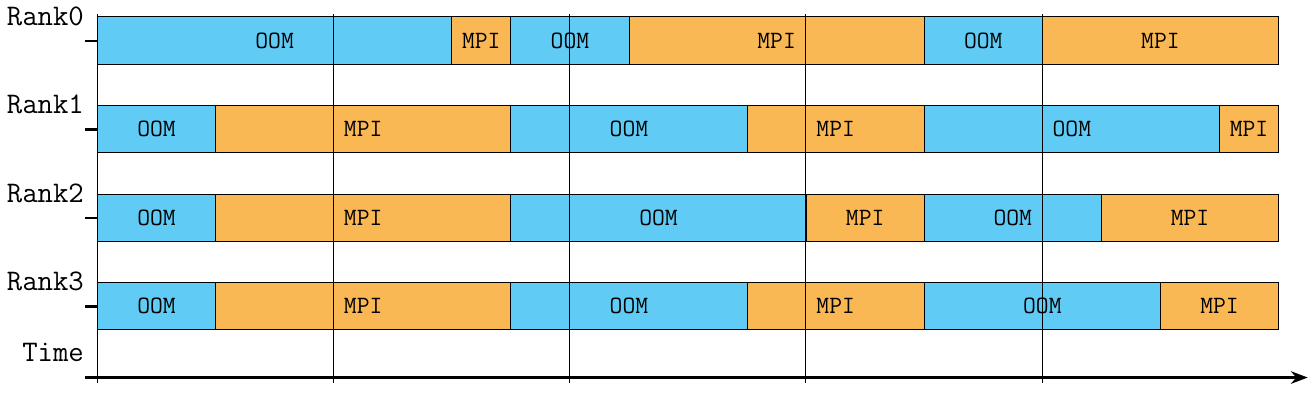}}
\end{minipage}
\begin{minipage}[c]{\linewidth}
  \subfloat[Discretised metrics calculation framework uses available critical path timers and interpolates all timers on discretisation window boundaries.\label{fig:disc-frameworkflow}]
  {\includegraphics[width=0.85\linewidth]{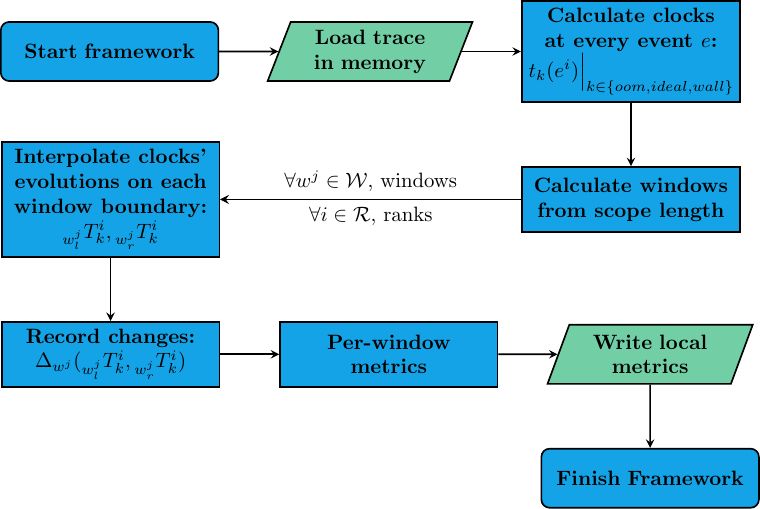}}
\end{minipage}
\caption{Timeline discretisation overview and discretised metrics calculation framework.}
\label{fig:discretisation}
\end{center}
\end{figure}
This framework calculates the change in the OOM times and the critical path length at every MPI event across all ranks in a trace.
This exercise enables local metric calculation at any interval by interpolating the timers' increases across the discretisation boundaries.
Metrics derived from timer values, aggregated across the entire execution, match those obtained with existing methods and tools.
This fact is used to validate the tool.

The time evolution of the metrics is calculated by dividing the timeline into repeating windows of small durations.
After that, the events are iterated to calculate changes in OOM elapsed time and local critical path length between two discretisation points over a time window.
Even though the transfer efficiency monitor~\cite{haldar24} cannot calculate load imbalance or serialisation, because it involves only one rank, it reveals the volatility of the metric values for consecutive repeated OOM/MPI duration pairs.
It showed that the scope of transfer efficiency calculation extends at least to an OOM region and the following MPI region of one rank.
Extending from there, each window should have multiple MPI events for all ranks to accurately reflect the meaningful bottlenecks.
Fig.~\ref{fig:disc-cartoon} presents the discretisation of the timeline of fixed length.
The local metrics are calculated for each discretised region separately (Fig.~\ref{fig:disc-frameworkflow}).

To facilitate this, three separate clocks are conceptualised that are initialised at the beginning of trace analysis and evolve according to the rules in Table~\ref{tab:clocks-evolve}.
Fig.~\ref{fig:discretisation} shows a rank's timeline of an OOM region followed by an MPI region resolved into wait and transfer durations via critical path analysis.
For each event point in Fig.~\ref{fig:discretisation}, the evolution of the clocks is described in Table~\ref{tab:clocks-evolve}.
\begin{figure}[tp]
\begin{center}
\includegraphics[width=.95\linewidth]{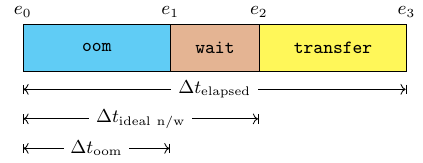}
\caption{The event points after critical path analysis resolve durations.}
\label{fig:discretisation}
\end{center}
\end{figure}

\begin{table}[tp]
\begin{center}
\caption{Various clocks run during different durations.}
\begin{tabular}{|r|l|l|l|l|}
  \hline
  { Clock } & { Value at $e_0$ } & { Value at $e_1$ } & { Value at $e_2$ } & { Value at $e_3$ }\\
  \hline
  {$T_\mathrm{elapsed}$} & $t_\mathrm{elapsed}^0$ & $t_\mathrm{elapsed}^0+ \Delta t_\mathrm{oom}$ & $t_\mathrm{elapsed}^0+ \Delta t_\mathrm{ideal~n/w}$ & $t_\mathrm{elapsed}^0+ \Delta t_\mathrm{elapsed}$\\
  {$T_\mathrm{oom}$} & $t_\mathrm{oom}^0$ & $t_\mathrm{oom}^0+ \Delta t_\mathrm{oom}$ & $t_\mathrm{oom}^0+ \Delta t_\mathrm{oom}$ & $t_\mathrm{oom}^0+ \Delta t_\mathrm{oom}$\\
  {$T_\mathrm{ideal~n/w}$} & $t_\mathrm{ideal~n/w}^0$ & $t_\mathrm{ideal~n/w}^0+ \Delta t_\mathrm{oom}$ & $t_\mathrm{ideal~n/w}^0+ \Delta t_\mathrm{ideal~n/w}$ & $t_\mathrm{ideal~n/w}^0+ \Delta t_\mathrm{ideal~n/w}$\\
  \hline
\end{tabular}
\label{tab:clocks-evolve}
\end{center}
\end{table}

\begin{figure}[tp]
 \begin{center}
  \begin{minipage}[]{\linewidth}
   \hspace{0pt}
  \subfloat[Part of a rank's timeline with resolved wait/transfer and a window boundary.\label{fig:evolve-timeline}]
  {\includegraphics[width=0.95\linewidth]{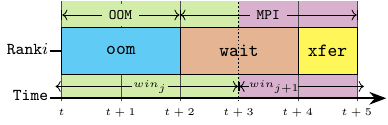}}
  \end{minipage}
  \begin{minipage}[]{\linewidth}
   \hspace{28pt}
  \subfloat[Evolution of the clocks across the above timeline.\label{fig:evolve-clock}]
  {\includegraphics[width=0.81\linewidth]{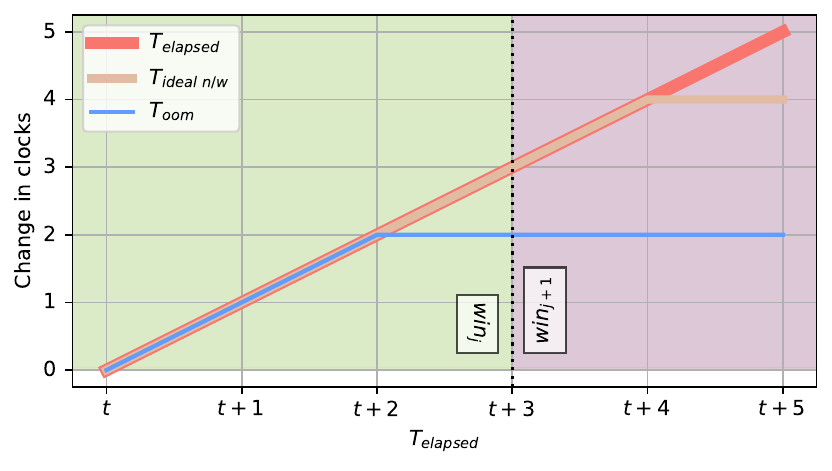}}
  \end{minipage}
  \begin{minipage}[]{\linewidth}
   \hspace{28pt}
  \subfloat[Contribution of the clocks' evolution divided into two windows due to wait-then-transfer model.\label{fig:evolve-clock-sep}]
  {\includegraphics[width=0.81\linewidth]{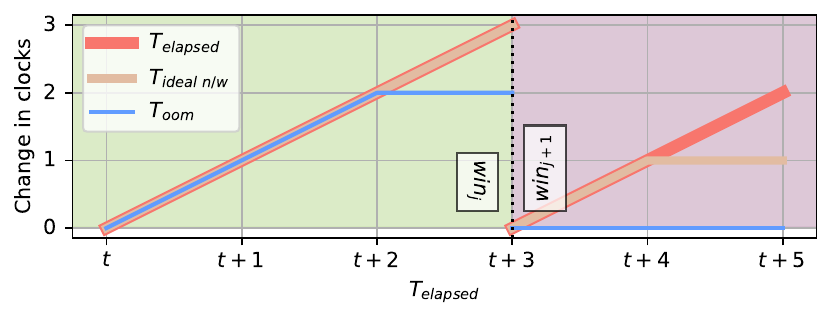}}
  \end{minipage}
 \end{center}
 \caption{A compute followed by an MPI region with a discretisation boundary through it: $\Delta T_{oom}$ only contributes in the first window; $\Delta T_{critical}$ and $\Delta T_{elapsed}$ contribute to both windows as per the model.}
 \label{fig:discretisation-boundary}
\end{figure}

Fig.~\ref{fig:evolve-timeline} has an OOM region and an MPI region of a rank, and the discretisation boundary is through the MPI duration.
Such situations are expected since different ranks perform different tasks simultaneously.
This work models time within the MPI runtime as \textit{wait-then-transfer}, where all wait durations from the critical path analysis are executed first, followed by the transfer durations.
After that, a special scheme is applied that divides the evolution of the clocks adhering to the wait-then-transfer model at the boundaries.
In the figure, the OOM and the critical path timers account for $40\%$ and $80\%$ of the total shown duration, respectively.
With the model, after $40\%$ of the time in the timeline, the $\Delta t_{oom}$ remains constant.
Similarly, the $\Delta t_{critical}$ remains constant after $80\%$ of the durations.
This model is realistic for OOM timers, but may not be exact for ideal-network timers in MPI calls that finish many messages, for example, in \texttt{MPI\_Waitall} calls.

Another modelling approach would be to increase the timers on a slope less than $1$ so they reach their final values at the end.
The latter approach is not explored because it does not accurately model any timer.
Notably, the elapsed timer always has a slope of $1$.

The critical paths are calculated by reconstructing the ideal-network timers from recorded traces.
The work extends the concept of Lamport logical clocks~\cite{lamport78} to use elapsed time since the last MPI exit rather than the original incremental counters.
The ideal network timers stop at the MPI entry points and do not run within the MPI runtime.
The distributed MPI event ordering is maintained by performing a compare-and-swap at the receiving side of each MPI call with the timer value from the sender.
Since the extended Lamport logical clocks maintain MPI's distributed operation order, this reconstructs the critical path timers.
From the MPI perspective, the Lamport timer pairs are:
\begin{itemize}
  \item when a message originates: the beginning of the send directive (such as \texttt{MPI\_Isend} or \texttt{MPI\_Start}); and
  \item when a message terminates: the end of the corresponding receive directive (such as \texttt{MPI\_Wait} or \texttt{MPI\_Recv}).
\end{itemize}
The end-of-wait timestamps ($e_2$ in Fig.~\ref{fig:discretisation}) in MPI calls are calculated by replaying MPI events from a trace and minimising MPI durations as much as possible, without violating the described causality of point-to-point messages.
The remaining MPI durations are solely spent waiting, as they cannot overlap with their partnering ranks for that message.
Existing tools calculate the critical path length for a given duration (usually the entire execution duration) using mathematically equivalent techniques.

\subsection{Effects of discretisation parameters}
Unlike the OOM timers, the ideal-network timer of one rank may increase more than the elapsed time between two events because ranks may become part of the critical path, defined as the rank with the maximum value of the critical path length at that point, as the application proceeds.
To maintain metric stability, the difference between the highest critical path lengths at the two boundaries of a time window is taken into account.
The super-elapsed increase in critical path lengths occurs more often for small windows, as critical paths cannot propagate among all MPI ranks in such cases.

When the window lengths are too short to capture any event, the local metrics provide little insight; they consist only of OOM or MPI durations.
The transfer efficiency monitors~\cite{haldar24} showed that the minimum scope of transfer efficiency is at least an OOM region and the following MPI region of one rank.
This work extends this policy to all ranks in any window for serialisation and computation-load balance calculation by adaptively merging consecutive windows to capture at least a threshold number of events.
This policy dictates that at least three trace events are required to accurately represent the localised performance, not artefacts of small windows.
With this logic, the first and the second windows merge in Fig.~\ref{fig:disc-cartoon} to obtain the minimum number of events.
So do the next two windows.

One practical example is when tracing is turned off for a relatively long time, and no events occur in that region.
However, maintaining the minimum threshold of three events may result in partial influence from neighbouring windows, since the timers increase at a nominal rate or remain unchanged.
In Fig.~\ref{fig:discretisation-boundary}, the $\Delta t_{oom}$ is absent in the second window and cannot contribute to load balance or serialisation calculation.
So, more consecutive durations are desirable for the least eventful rank.
This work performed most evaluations with a threshold of $8$ events or at least $3$ consecutive pairs of compute and MPI regions, unless otherwise noted.

\section{Implementation of Discretised Metric Computation}
\label{sec:tool-des}
This work implements the framework from section~\ref{sec:discrete-framework} in a tool called \texttt{ClockTalk}~\cite{clktk}, which supports Paraver format traces of MPI applications.
The discretisation length is a user-supplied value for the tool.
The tool processes Paraver traces of MPI applications and calculates the fundamental performance factors for the entire duration.
It reads and stores all MPI events and their timestamps from a trace in memory, then replays the trace to calculate the OOM and ideal-network timers using extended Lamport clocks at each event.

The ideal network calculation implements an eager limit for point-to-point communications, requiring the sender to wait for larger messages.
It supports multiple communicators for collective communication, which increases the memory requirements for replay.
After the replay, the calculated OOM timers and critical path lengths are stored at each event.
The trace events can be parsed per process, making the access convenient for the discretisation framework.
If the minimum threshold of MPI events is not reached for at least one rank, the length is doubled until every rank contains at least that many events.

The framework iterates through the events to update the timers for every rank in a window.
The timer values are interpolated on the boundary after the window is crossed in all ranks.
Changes in the interpolated timers calculate the metrics in that window.
For the discretisation boundaries throughout OOM regions, both timers run as usual.
The discretisation boundary through MPI regions requires the treatment described in Section~\ref{sec:discrete-framework}.

\subsubsection{Incorrect Message Matching}
The tracing tool's message-matching module may incorrectly match messages.
In one case study, a receive completion, followed by a blocking collective on the world communicator, and then the corresponding send was observed.
The tool detects such faulty pairs and treats them as local MPI operations, allowing the analysis to proceed meaningfully.
This feature is a clear improvement over current tool practices (discussed in Section~\ref{subsec:real-apps}).

\section{Case Studies}
\label{sec:eval}
Traces from a synthetic benchmark application and two real-world applications: a 3-D finite-difference code (LaMEM) and a molecular dynamics code (ls1-MarDyn) are used for case studies.
The benchmark and ls1-MarDyn were run on 2 and 16 AMD EPYC 7742 nodes at 2.25 GHz with $256~GB$ of memory per node on the Hawk system at the High Performance Computing Center Stuttgart (HLRS).
LaMEM was run on 1 or 2 AMD EPYC 7763 nodes at 2.45 GHz with $256~GB$ of memory per node on the LUMI-C system at the Finnish IT Center for Science (CSC).
From the Paraver traces, the existing tool (BSC toolchain, which uses the Dimemas~\cite{dimemas97} for the critical path calculation) calculated the metrics, and this work's tool calculated the localised metrics.
The final metrics values from both tools are compared for validation, and agreement is observed when the Dimemas runs were successful.

\subsection{Synthetic Benchmark Application}
This work modified a 2-D heat-equation solver with a 5-point stencil that implements successive over-relaxation and ran it with 256 MPI processes.
The modified solver exhibits four regions with varying amounts of computation and data transfer, depending on the amount of computation at different times.
The four regions exhibit very different performance characteristics, as evident from the method applied to a recorded trace.
The behaviours mimic situations exhibiting imbalanced compute loads and/or running on a slower network.
The local metrics differ from the aggregated metrics computed by the BSC toolchain for most of the timeline.
The region between 350 and 520 seconds suffers from serialisation, whereas the region from 520 seconds to the end suffers from inefficient transfers.
These effects are readily identified using the presented method.
\begin{figure}[tp]
\begin{minipage}[]{.75\linewidth}
  \includegraphics[width=\linewidth]{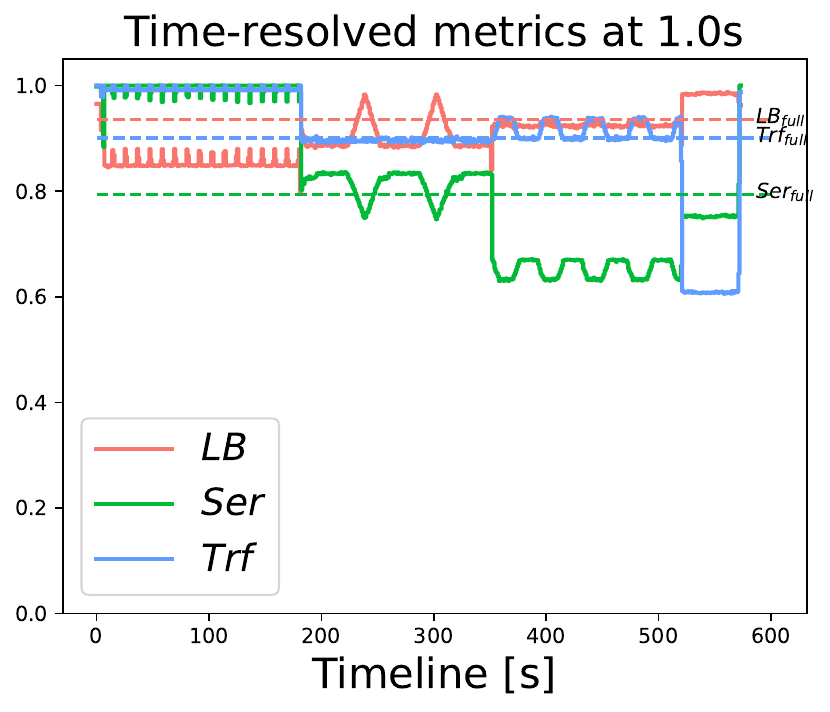}
\end{minipage}
\begin{minipage}[]{.242\linewidth}
  \includegraphics[width=\linewidth]{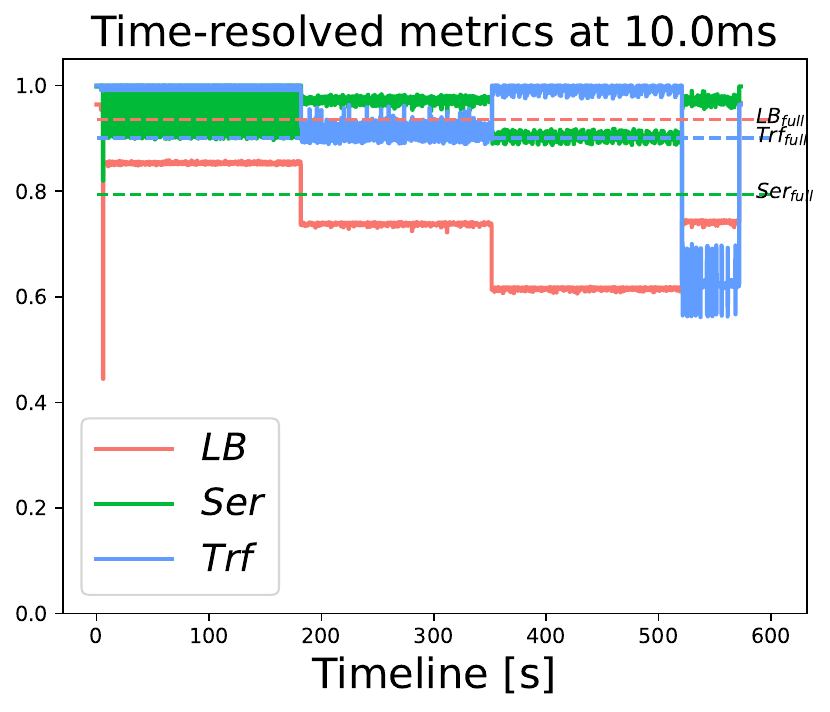}
  \includegraphics[width=\linewidth]{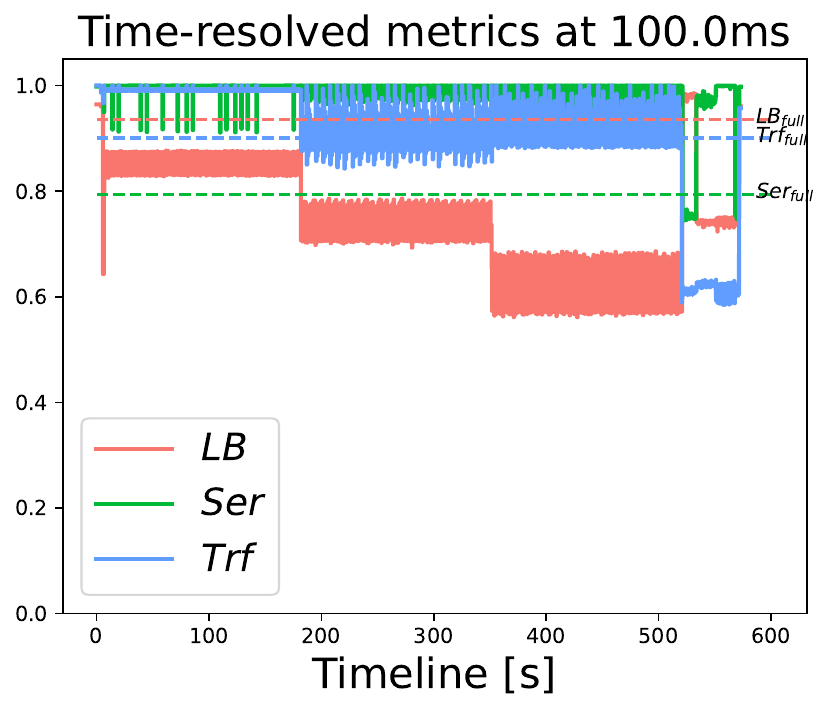}
  \includegraphics[width=\linewidth]{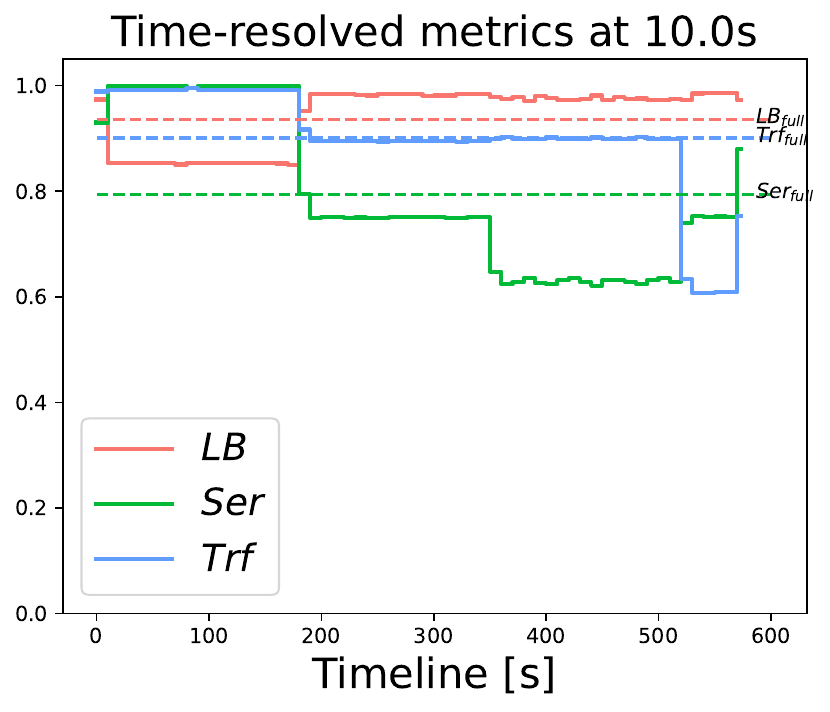}
\end{minipage}
\caption{2-D stencil benchmark: Time-resolved metric plots with different window sizes. Smaller windows reveal transient bottlenecks (spikes) that are averaged out in longer windows, demonstrating the benefit of fine-grained temporal analysis. The dashed lines are the metrics for the entire execution duration.}
\label{fig:sor-method}
\end{figure}

\begin{figure}[tp]
 \begin{center}
  \begin{minipage}[]{.242\linewidth}
    \raisebox{0ex}{\includegraphics[width=\linewidth]{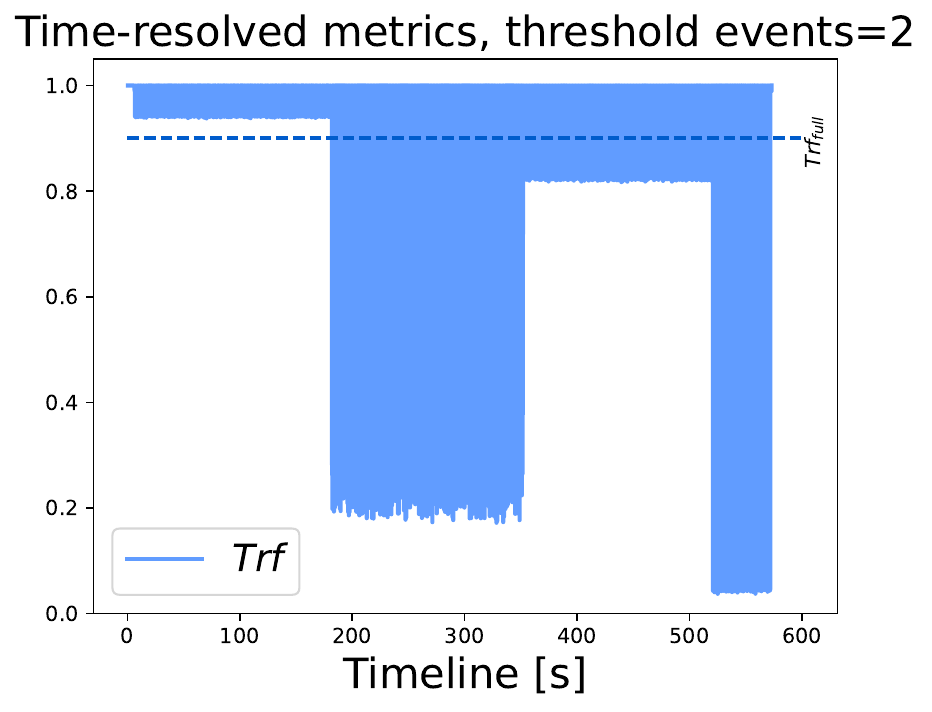}}
  \end{minipage}
  \begin{minipage}[]{0.242\linewidth}
    \raisebox{0ex}{\includegraphics[width=\linewidth]{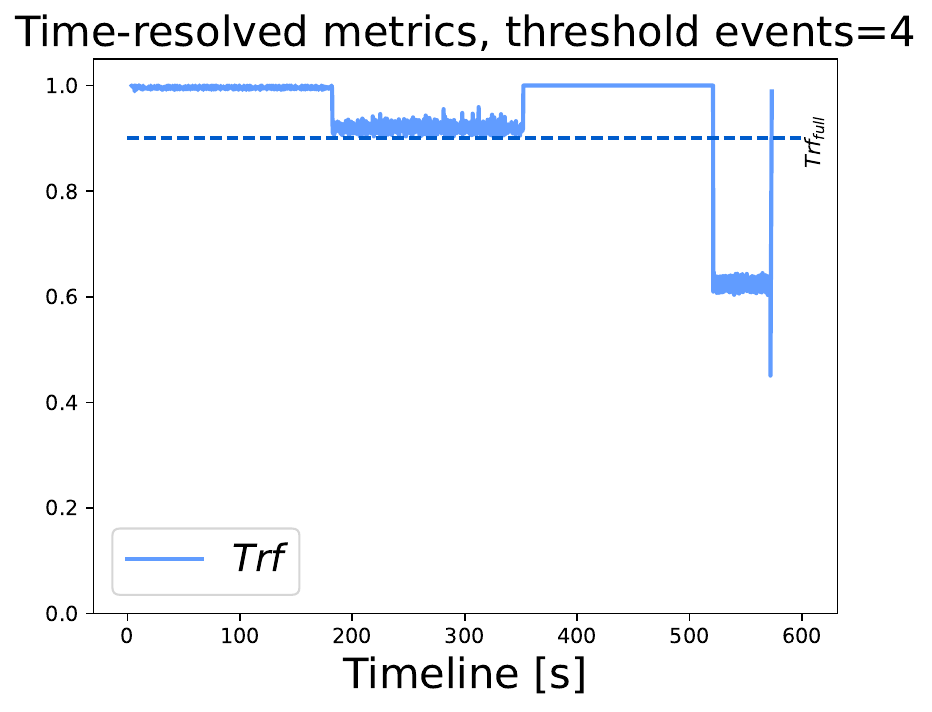}}
  \end{minipage}
  \begin{minipage}[]{0.242\linewidth}
    \raisebox{0ex}{\includegraphics[width=\linewidth]{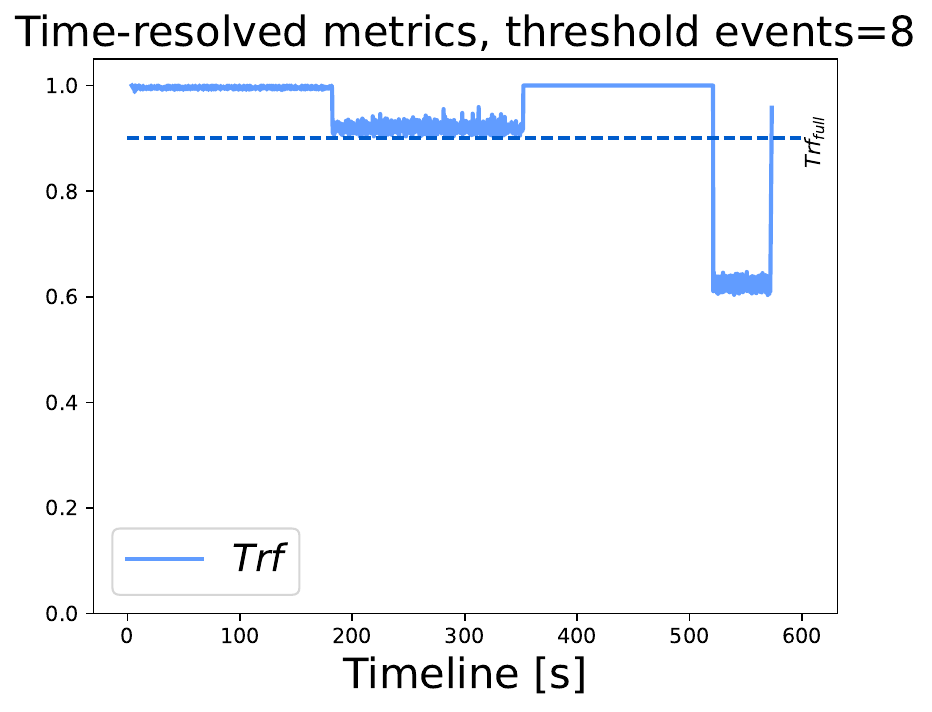}}
  \end{minipage}
  \begin{minipage}[]{0.242\linewidth}
    \raisebox{0ex}{\includegraphics[width=\linewidth]{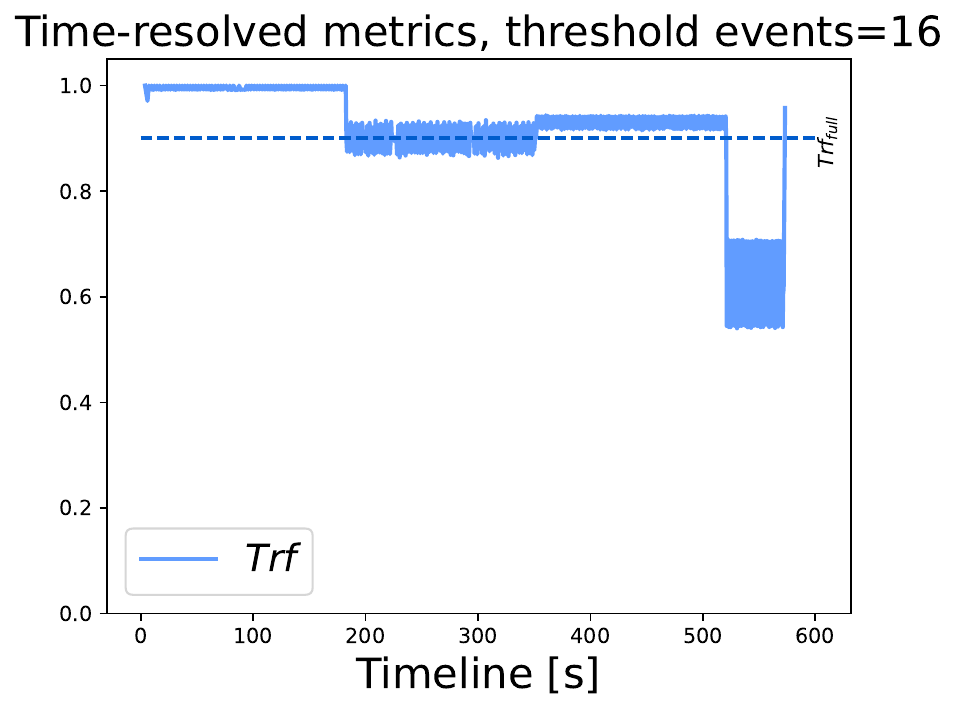}}
  \end{minipage}
 \end{center}
 \caption{2-D stencil app: At least three events are needed to avoid the discretisation artefacts. The dashed line is from the BSC toolchain.}
 \label{fig:sor-nevts}
\end{figure}

Fig.~\ref{fig:sor-method} shows the effect of four different window sizes on the tool's output.
The $10s$ window exhibits the most general characteristics the analysts would see with trace visualisation before further analysis.
The smaller windows reporting the serialisation as imbalanced compute aligns with the work by Huck et al.~\cite{huck10}.
Even though the overall transfer efficiency is close to $90\%$, the poor transfer can be separately investigated.
Such insights are absent in the standard aggregated metrics and can only be qualitatively inferred from a trace visualisation.
Fig.~\ref{fig:sor-nevts} shows the effect of the minimum event threshold on the transfer efficiency.
The other two metrics also exhibit similar behaviour.

OTF2 traces using Score-P~\cite{scorep12} were obtained for the benchmark, and used Scalasca/Scout for a more exhaustive comparison.
It calculates per-function metrics but cannot resolve characteristics arising from the changed computational load in the same function due to the use case.
The obtained values, however, agree with those of the BSC toolchain and the ClockTalk tool.

\subsection{Real-World Applications}
\label{subsec:real-apps}
\subsubsection{LaMEM}\cite{lamem} is a PETSc-based 3-D finite difference solver.
\begin{figure}
\centering
\subfloat[64 MPI processes: the dashed lines are by the BSC toolchain\label{fig:lamem-064}]{%
  \begin{minipage}[]{.242\linewidth}
    \raisebox{0ex}{\includegraphics[width=\linewidth]{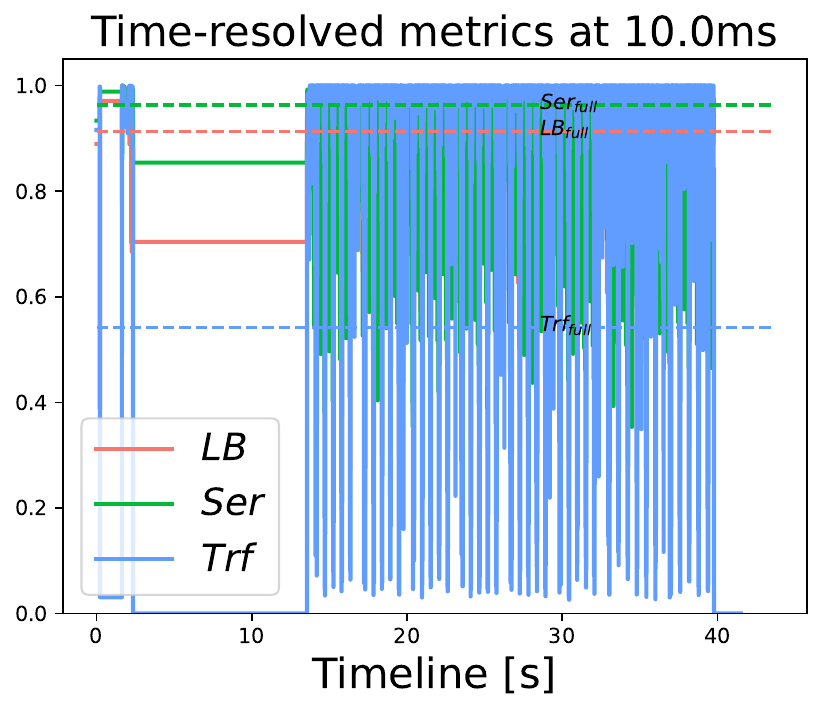}}
  \end{minipage}
  \begin{minipage}[]{0.242\linewidth}
    \raisebox{0ex}{\includegraphics[width=\linewidth]{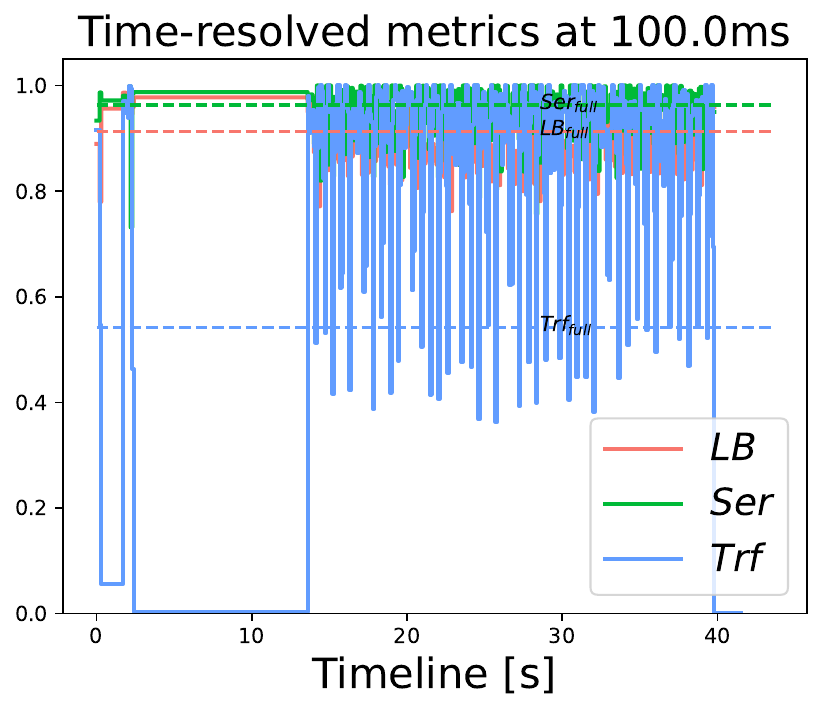}}
  \end{minipage}
  \begin{minipage}[]{0.242\linewidth}
    \raisebox{0ex}{\includegraphics[width=\linewidth]{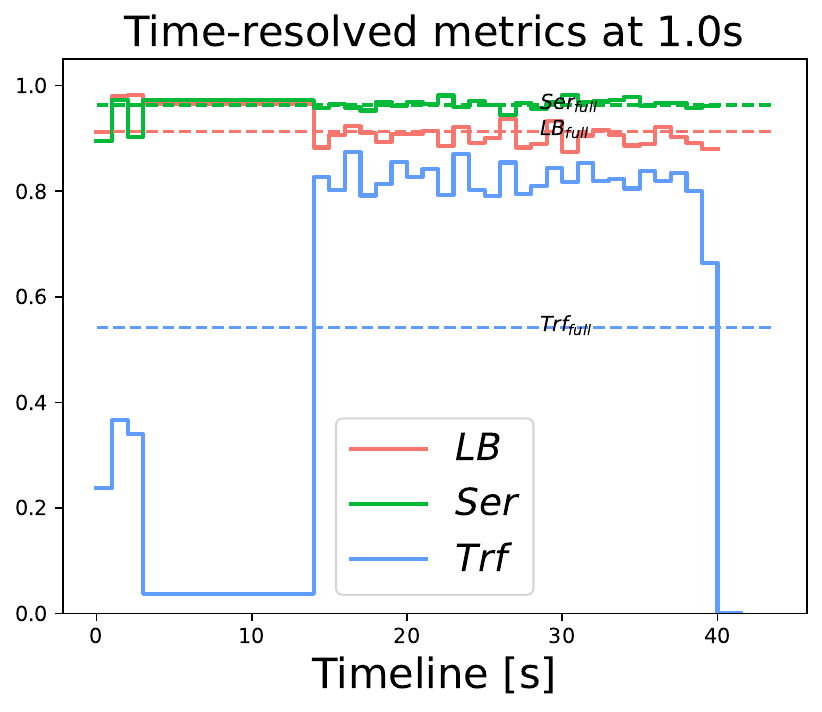}}
  \end{minipage}
  \begin{minipage}[]{0.242\linewidth}
    \raisebox{0ex}{\includegraphics[width=\linewidth]{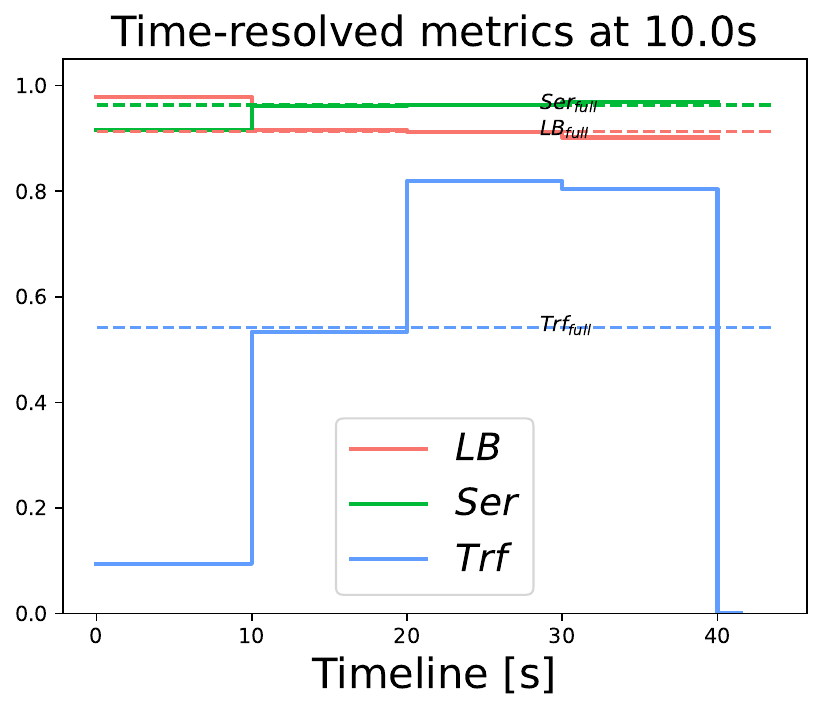}}
  \end{minipage}
}
\hfil
\subfloat[128 MPI processes: the dashed lines are aggregated results of ClockTalk\label{fig:lamem-128}]{%
  \begin{minipage}[]{.242\linewidth}
    \raisebox{0ex}{\includegraphics[width=\linewidth]{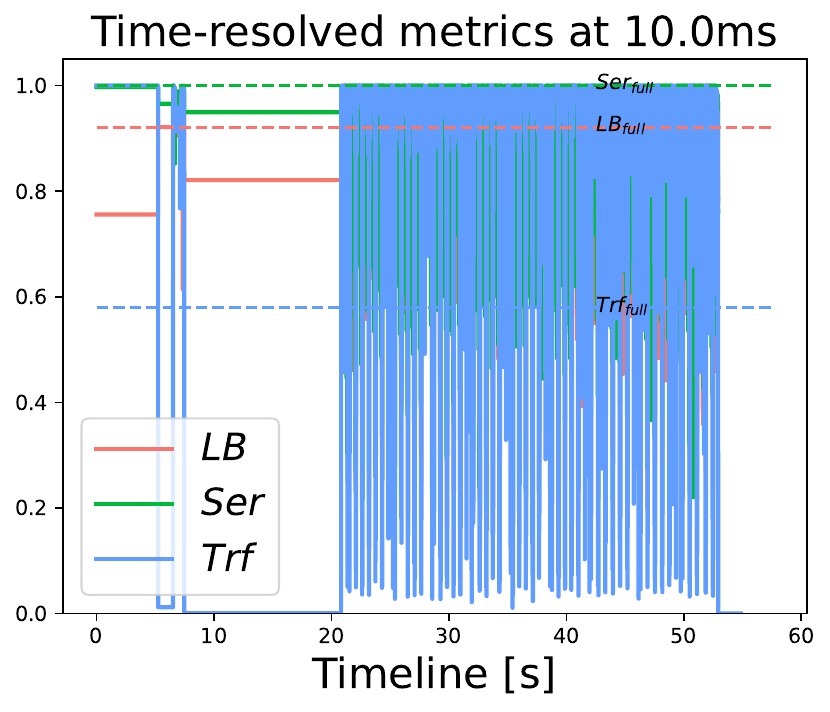}}
  \end{minipage}
  \begin{minipage}[]{0.242\linewidth}
    \raisebox{0ex}{\includegraphics[width=\linewidth]{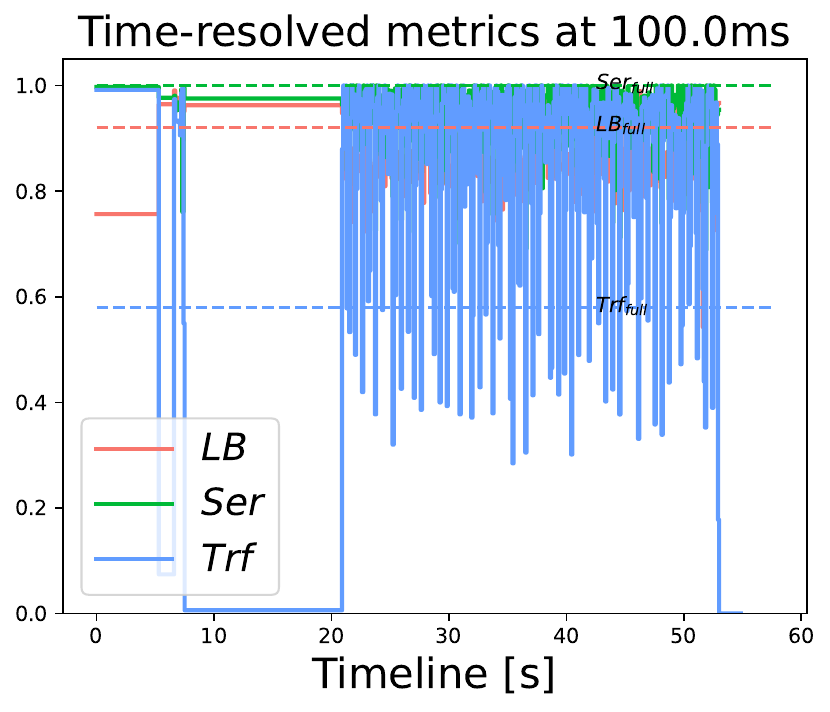}}
  \end{minipage}
  \begin{minipage}[]{0.242\linewidth}
    \raisebox{0ex}{\includegraphics[width=\linewidth]{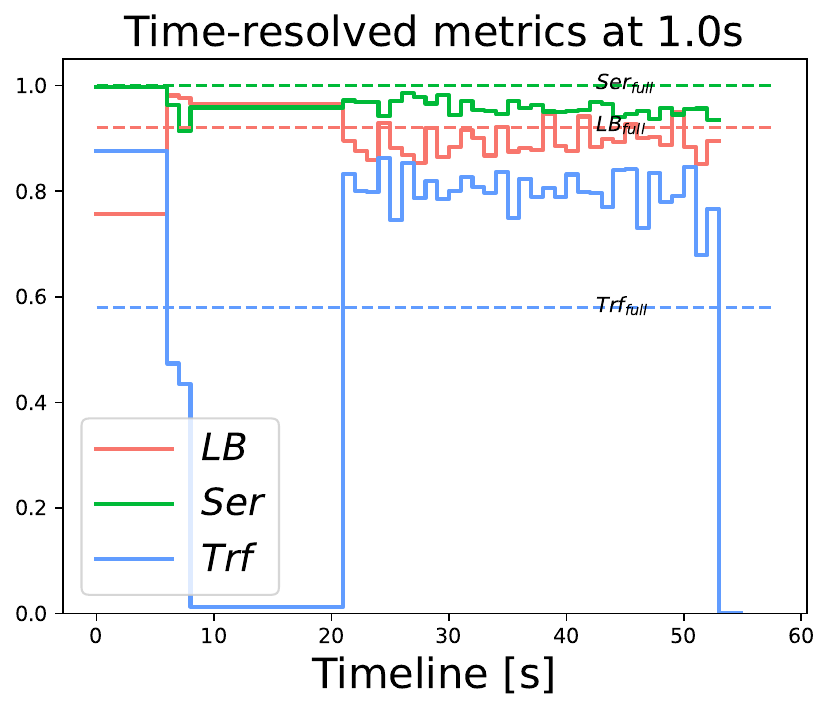}}
  \end{minipage}
  \begin{minipage}[]{0.242\linewidth}
    \raisebox{0ex}{\includegraphics[width=\linewidth]{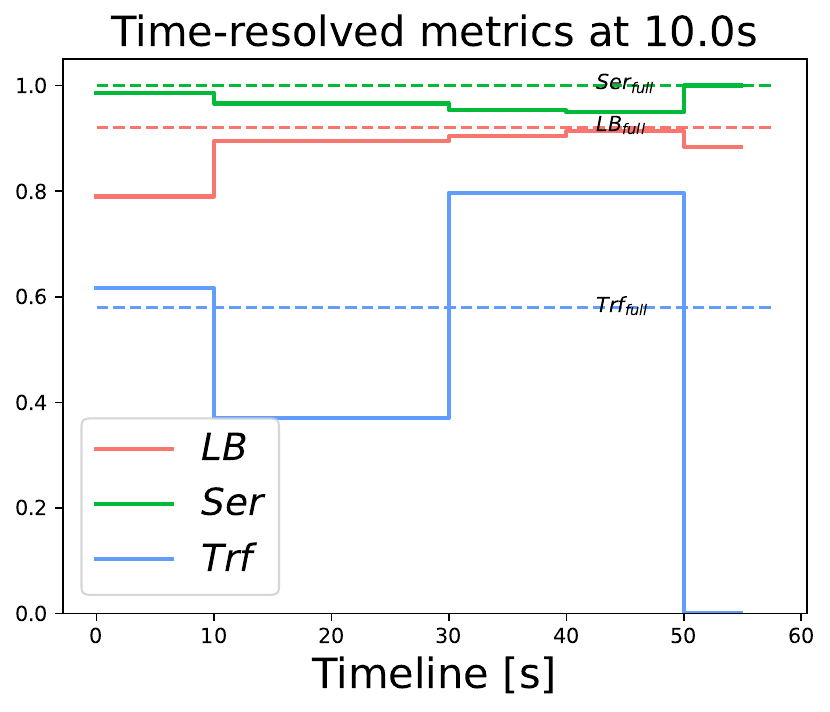}}
  \end{minipage}
}
\caption{LaMEM: time-resolved metrics at various resolutions. The aggregated metrics for the 128 processes are from the ClockTalk tool since the BSC toolchain failed.}
\label{fig:lamem-method}
\end{figure}
It was run with 64 and 128 MPI processes for performance audits.
These traces contain user instrumentation to trace only intended regions.
The absence of data points in the figures after the initial values, until about $13s$ and $21s$ for Fig.~\ref{fig:lamem-064} and Fig.~\ref{fig:lamem-128}, respectively, indicates that tracing was turned off in those regions.
The BSC toolchain failed for the latter use case, whose size is $3.8~GB$.
ClockTalk ran and provided both time-resolved and aggregated metrics.

ClockTalk reported wrongly matched send-receive pairs in the trace, resulting in the receive finishing before the corresponding send starts.
The tool identified and ignored them, enabling the audit to proceed.
This fix is a clear improvement over the state of the art, as even when results from such pairs are ignored, the overall metric trend can be retained with the tool.
Direct analyses of the trace without visualising were possible with the tool by identifying and isolating specific regions (Fig.~\ref{fig:lamem-method}).
The trends were verified by opening the traces in Paraver for visual inspection.
Due to its small size, Paraver visual inspection would also allow qualitative visual analysis.
%
%
\subsubsection{ls1-MarDyn}~\cite{ls1} is a massively parallel Molecular Dynamics (MD) code for large systems, targeting thermodynamics and nanofluidics.
A spinodal decomposition use case with $2048$ MPI processes was executed for around $10$ minutes.
\begin{figure}[tp]
 \begin{center}
  \begin{minipage}[]{.242\linewidth}
    \raisebox{0ex}{\includegraphics[width=\linewidth]{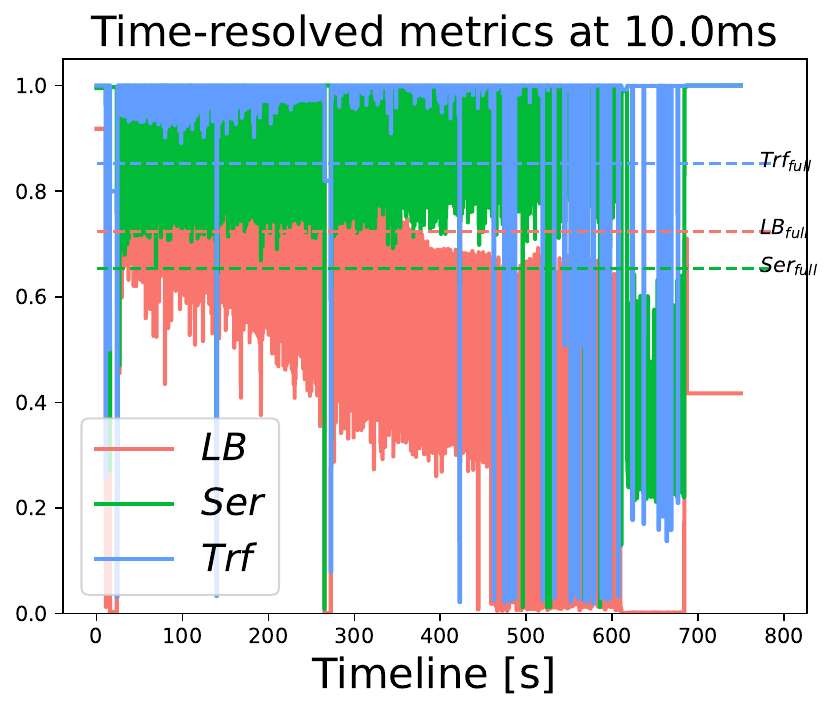}}
  \end{minipage}
  \begin{minipage}[]{0.242\linewidth}
    \raisebox{0ex}{\includegraphics[width=\linewidth]{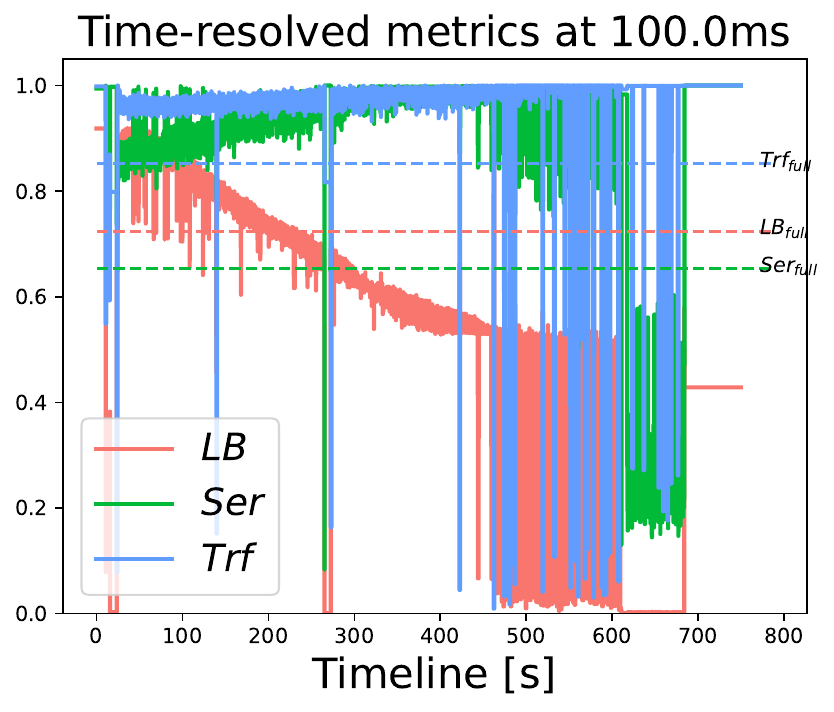}}
  \end{minipage}
  \begin{minipage}[]{0.242\linewidth}
    \raisebox{0ex}{\includegraphics[width=\linewidth]{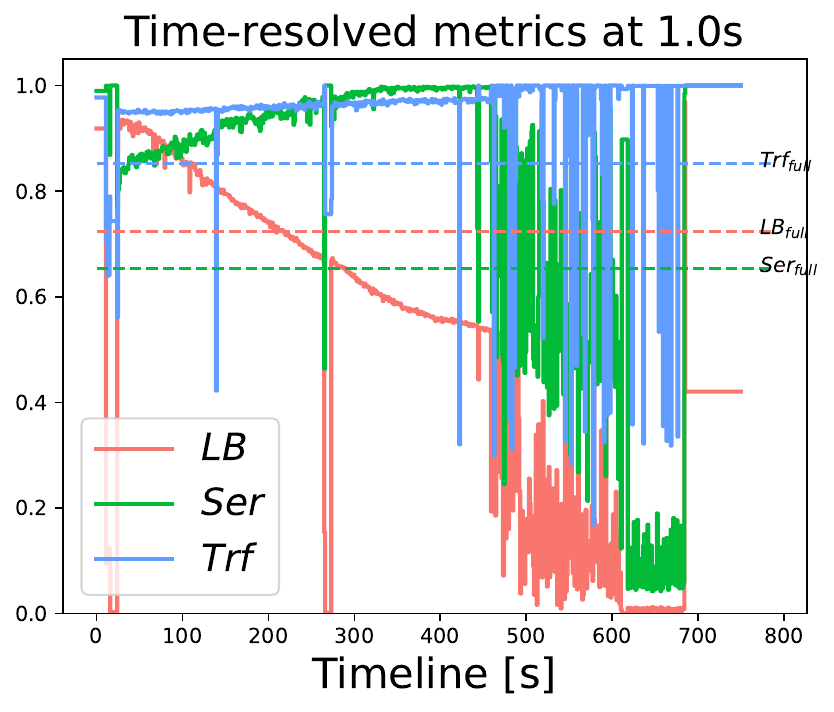}}
  \end{minipage}
  \begin{minipage}[]{0.242\linewidth}
    \raisebox{0ex}{\includegraphics[width=\linewidth]{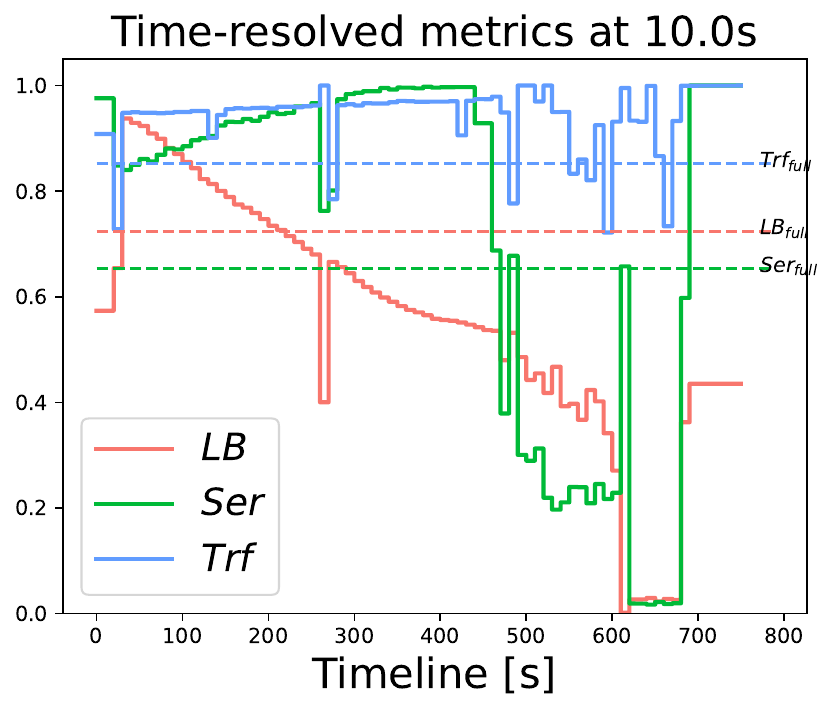}}
  \end{minipage}
 \end{center}
 \caption{ls1-MarDyn: Time-resolved metrics at various resolutions. The dashed lines are aggregated metrics calculated using the ClockTalk tool.}
 \label{fig:mardyn-method}
\end{figure}
Even though the large trace volume ($465 GB$) is unsuitable for visualisation or analysis with the BSC toolchain, ClockTalk generated the time-resolved performance metrics.
This use case benefited from the performance metrics visualisation, which is designed to collect more molecules in specific regions as iterations progress.
Such a visualisation was not possible before.

Fig.~\ref{fig:mardyn-method} shows that the compute-load imbalance increases until the end.
The other metrics behave differently, such as serialisation efficiency, which increases from an initial $85 \%$ to close to $100 \%$.
It also captured the I/O event before $300s$ and examined its impact on local performance.
Without the presented method, trace segmentation would be performed without accounting for different regions, necessitating separate analyses.
The calculation cut-off was set to $6000s$ for this case.
The trace buffers overflowed after about $440s$, and every rank started flushing the buffers whenever an overflow occurred.
The overflow resulted in metric values that did not represent the application, but rather a tracing artefact.

\subsection{Analysis duration}
\begin{table}
\begin{center}
\caption{Execution times of BSC toolchain and ClockTalk}
\label{tab:tool-times}
\begin{tabular}{|r|r|r|r|r|}
\hline
{ Use case } & { Trace }     & Application & { BSC } & { ClockTalk }\\
             & { size [GB] } & ranks       & { toolchain [s] } & { tool [s] }\\
\hline
{Bench app.} & $1.3$           & $256$         & $554.50$       & $9.06$ \\
{LaMEM-1}    & $1.4$           & $64$          & $905.21$       & $17.38$\\
{LaMEM-2}    & $3.8$           & $128$         &$1459.15$       & $44.66$\\
{ls1-MarDyn} & $465.0$         & $2048$        & Time out     & $2950.47$\\
\hline
\end{tabular}
\end{center}
\end{table}
Table~\ref{tab:tool-times} compares execution times of the tools on a single node of the Hawk system.
By avoiding the I/O (as is the case with the BSC toolchain), ClockTalk executes several times faster.

\section{Related Work}
\label{sec:rel-work}
Gonzalez et al.~\cite{clustering09} proposed a density-based clustering strategy to isolate computational phases in MPI applications adhering to the single-process, multiple-data (SPMD) paradigm.
Gonzales et al.~\cite{sequence13} further improved this method by combining it with the Multiple Sequence Analysis (MSA) method.
Casas et al. ~ \cite{wavelet08} and Casas et al. ~ \cite{spectral10} developed a generalised framework to apply wavelet transformation techniques to identify repeating execution structures and determine the statistically most representative area of these structures, which can be readily visualised and analysed.
Scalasca~\cite{scalasca10} processes the trace at the invocation of \texttt{MPI\_Finalize} to perform root-cause analysis~\cite{scalascaroot16} and stores it in the trace to show as hotspots.
Dimemas~\cite{dimemas97} is used in the BSC toolchain to simulate a special network with no latency and infinite bandwidth to calculate the critical path length.
Alawneh et al.~\cite{Alawneh2016SegmentingSystems} recognise hierarchical patterns of MPI communications and exploit differences in their information entropy to segment trace regions.
Later modifications by Alawneh et al.~\cite{Alawneh2022LocatingTraces} identify communication patterns with more spread-out durations than others.
Lopez et al.~\cite{talp21} track the load balance metric online to detect and manage the computational load balance.
Protze et al.~\cite{otf22} calculate the performance factors with an extra MPI call for each non-local MPI function to exchange timers.
mpiP~\cite{mpip} and Score-P profiler provide partial metrics for the entire execution.


\section {Summary and Outlook}
\label{sec:summary}
This work described the mathematical formulation of the time-resolved window in the timeline and how the tool can generate important performance insights from traces, especially when timeline visualisation is difficult or impractical.
The method's applicability to various synthetic and real-world applications was evaluated, along with the effect of discretisation lengths on the results.
The tool shows changes in metric trends, but not all such changes reflect actual performance bottlenecks.
Analysts must interpret them within the context of the application.
The case studies showed that the visual cues react accurately and reliably to the evolving nature of the application's instantaneous performance.
The tool's consistency in resolving the metrics is demonstrated by comparing the values obtained from the standard tool.
The tool also identifies trace mistakes and can recover when the existing method fails.
This method will enable analysts to investigate larger traces more easily.
Currently, the tool works correctly with Paraver format traces from MPI-only applications.
In the future, the aim is to extend the tool to cover shared-memory paradigms and address the lack of probing of one-sided communication during tracing.

\begin{credits}
\subsubsection{\ackname} This work was partially funded by the German Federal Ministry of Research, Technology and Space (BMFTR) under grant agreement 16ME0584. The author is grateful to Jos\'e Gracia for providing valuable insights and access to the \texttt{LaMEM} trace files.

\subsubsection{\discintname}
\end{credits}
%
%
%
\bibliographystyle{splncs04}
\bibliography{refs.bib}
\end{document}